\pdfoutput=1
\documentclass[runningheads]{llncs}
\usepackage[T1]{fontenc}
\usepackage{amsmath, graphicx, hyperref, multirow, subfigure, caption, booktabs, array, makecell, bm}
\usepackage[hyphenbreaks]{breakurl}

\begin{document}
\title{Source-Filter-Based Generative Adversarial Neural Vocoder for High Fidelity Speech Synthesis}
%
%
\author{Ye-Xin Lu \and Yang Ai \and Zhen-Hua Ling}
%
\titlerunning{Source-Filter-Based Generative Adversarial Neural Vocoder}
%

\institute{National Engineering Research Center of Speech and Language Information Processing, \\University of Science and Technology of China, Hefei, P. R. China \\
\email{yxlu0102@mail.ustc.edu.cn},  \email{\{yangai, zhling\}@ustc.edu.cn}}

\maketitle              
\begin{abstract}

 This paper proposes a source-filter-based generative adversarial neural vocoder named SF-GAN, which achieves high-fidelity waveform generation from input acoustic features by introducing F0-based source excitation signals to a neural filter framework. The SF-GAN vocoder is composed of a source module and a resolution-wise conditional filter module and is trained based on generative adversarial strategies. The source module produces an excitation signal from the F0 information, then the resolution-wise convolutional filter module combines the excitation signal with processed acoustic features at various temporal resolutions and finally reconstructs the raw waveform. The experimental results show that our proposed SF-GAN vocoder outperforms the state-of-the-art HiFi-GAN and Fre-GAN in both analysis-synthesis (AS) and text-to-speech (TTS) tasks, and the synthesized speech quality of SF-GAN is comparable to the ground-truth audio.

\end{abstract}
\section{Introduction}
\label{sec:intro}

Speech synthesis, a technology that converts text information into human-like speech, has been actively studied by researchers to chase the high naturalness, intelligibility, and expressiveness of synthesized speech. In recent years, statistical parametric speech synthesis (SPSS) \cite{zen2009statistical} has become the dominant speech synthesis method due to its small system size, high robustness, and flexibility. SPSS generally consists of two stages: \textit{text to acoustic feature} and \textit{acoustic feature to speech}, which are implemented by acoustic models and vocoders, respectively.

For years, conventional vocoders such as STRAIGHT \cite{kawahara1999restructuring} and WORLD \cite{morise2016world} have been widely used in the SPSS framework. Because of their source-filter architecture, these vocoders tend to have good controllability of acoustic components. While subject to the signal-processing mechanism, some deficiencies exist, such as the loss of spectral details and phase information, which lead to the degradation of speech quality. 

Thanks to the developments of deep learning and neural networks, neural vocoders rapidly emerged and made great progress. Initially, autoregressive (AR) neural waveform generation models symbolized by WaveNet \cite{oord2016wavenet}, SampleRNN \cite{mehri2016samplernn}, and WaveRNN \cite{kalchbrenner2018efficient} were proposed to build neural vocoders and achieved breakthroughs in synthesized speech quality compared to the conventional methods. However, due to the AR structure, they struggle in inference efficiency when synthesizing high temporal resolution waveforms. Subsequently, to address this problem, knowledge-distilling-based models (e.g., Parallel WaveNet \cite{oord2017parallel}, and ClariNet \cite{ping2018clarinet}), flow-based models (e.g., WaveGlow \cite{prenger2018waveglow}, and WaveFlow \cite{ping2020waveflow}) were proposed. Although the inference efficiency has made significant progress, the computational complexity is still considerable, which limits their applications in resource-constrained scenarios.


Recently, researchers have been prone to pay more attention to waveform generation models without AR or flow-like structures. The neural source-filter (NSF) model \cite{wang2019neural} combines speech production mechanisms with neural networks and realizes the prediction of speech waveform from explicit F0 and mel-spectrogram. Generative adversarial networks (GANs) based models \cite{goodfellow2014generative,donahue2018adversarial,binkowski2019high,kumar2019melgan,yamamoto2020parallel,kong2020hifi}, simultaneously improved the synthesized speech quality with GANs and inference efficiency with parallelizable inference process. Among them, HiFi-GAN has achieved both high-fidelity and efficient speech synthesis. For the generative process, HiFi-GAN cascades multiple upsampling layers and multi-receptive field fusion (MRF) modules that consist of parallel residual blocks to gradually upsample the input mel-spectrogram to the temporal resolution of the final waveform while performing convolution operations. And for the discriminative process, HiFi-GAN adopts adversarial training with a multi-period discriminator (MPD) and a multi-scale discriminator (MSD) to ensure high-fidelity waveform generation. Based on it, Fre-GAN \cite{kim2021fre} further improved the speech quality in frequency space by adopting a resolution-connected generator and resolution-wise discriminators to capture various levels of spectral distributions over multiple frequency bands. However, due to their entirely data-driven manner, compared to source-filter-based vocoders, they tend to lack the controllability of speech components and robustness. Lately, there sprung some generative adversarial neural vocoders based on the source-filter model, such as quasi-periodic Parallel WaveGAN (QPPWG) \cite{wu2021quasi}, unified source-filter GAN (uSFGAN) \cite{yoneyama2021unified}, and harmonic plus noise uSFGAN (HN-uSFGAN) \cite{yoneyama2022unified}. Although they successfully achieved high-controllability speech generation, there still exists a gap in speech quality between synthesized and ground-truth audios.

To achieve high-fidelity speech synthesis, we propose SF-GAN, which integrates the source excitation into a GAN-based neural waveform model. Firstly, we design a source module through which the voiced and unvoiced segments of the excitation signal are generated from the upsampled F0 and Gaussian noise processed by a deep neural network (DNN), respectively. Secondly, we propose a resolution-wise conditional filter module to reconstruct speech waveform from the excitation signal and mel-spectrogram. Inspired by the layer-wise upsampling architecture in HiFi-GAN,  we subsample the excitation signal to various resolutions and subsequently condition them to each upsampling layer of HiFi-GAN. And then, we redesign the residual blocks in the MRF module by using parallel convolutions to combine the transformed excitation signals with the hierarchical intermediate features processed from mel-spectrogram. Additionally, to further apply our model to a text-to-speech (TTS) system, we design an F0 predictor to predict F0 from mel-spectrogram generated by an acoustic model. We use both the predicted F0 and mel-spectrogram as our model inputs. Both objective and subjective evaluations demonstrate that our proposed SF-GAN vocoder outperforms HiFi-GAN and Fre-GAN in the aspect of speech quality.

\begin{figure}[t!]
\includegraphics[width=\textwidth]{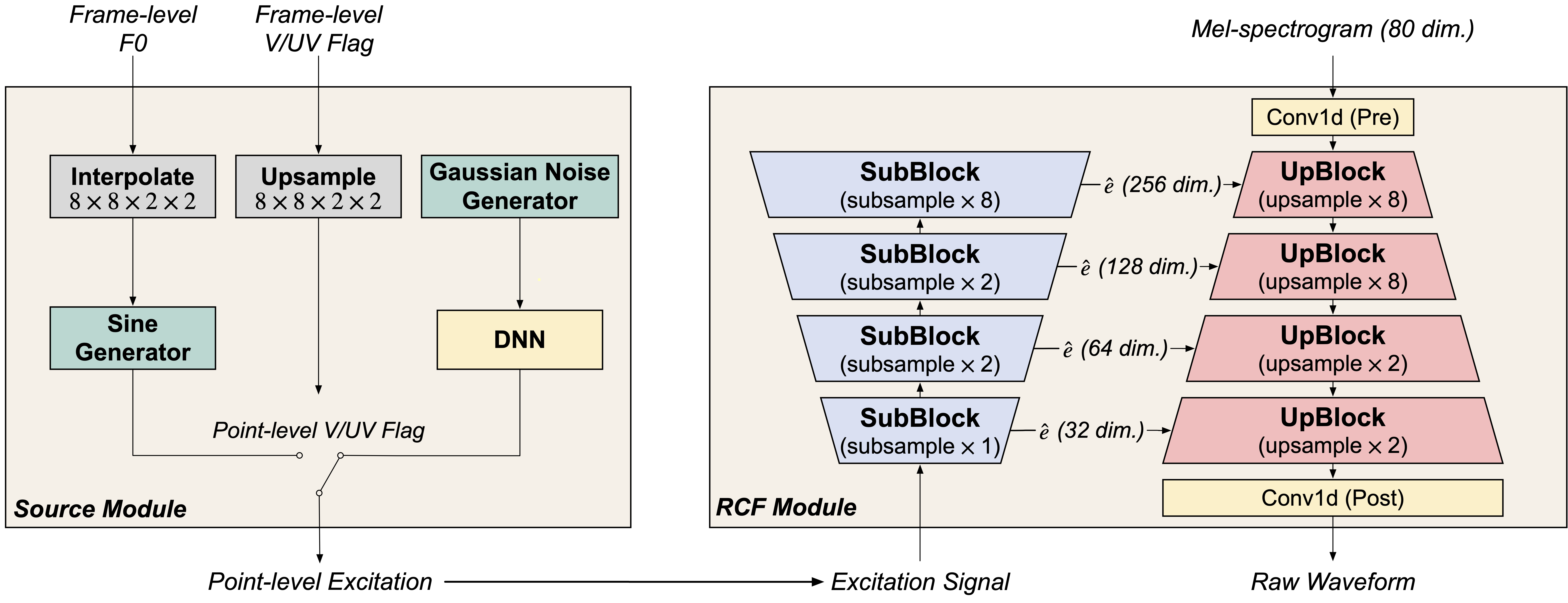}
\caption{The architecture of SF-GAN. In the source module, the V/UV flag is the abbreviation of the Voiced/Unvoiced flag. In the resolution-wise convolutional filter module, SubBlock denotes the subsampling residual block, while UpBlock denotes the concatenation of a transposed convolution and a parallel convolutional residual block.}\label{fig:1}
\end{figure}

\section{Proposed method}
\label{sec: proposed method}

\subsection{Overview}

In this paper, we propose the SF-GAN vocoder, which applies a source excitation access method to the HiFi-GAN architecture. As illustrated in Fig.~\ref{fig:1}, the proposed SF-GAN vocoder consists of a source module and a filter module, and it takes frame-level F0 and mel-spectrogram as inputs and outputs a speech waveform. The source module converts frame-level F0 into point-level excitation signal, while the filter module reconstructs raw waveform from this excitation along with the mel-spectrogram. Our proposed model adopts the GAN-based training strategy, and we use the MSD and MPD in HiFi-GAN to capture consecutive and periodic patterns.

\subsection{Source module}

Given the frame-level F0 sequence $\bm{f}_{1: L}$ as input, where $L$ denotes the number of frames, the source module firstly extracts the voiced/unvoiced (V/UV) flag sequence $\pmb{v}_{1: L}$ from it. Subsequently, $\bm{f}_{1: L}$ is interpolated $N = T/L$ times to match the temporal resolution of the raw waveform, where $T$ denotes the number of waveform sample points and $\bm{v}_{1: L}$ is also upsampled $N$ times by repeating V/UV flag values in each frame. The interpolated F0 sequence $\bm{f}_{1: T}$ with V/UV flags sequence $\bm{v}_{1: T}$ are converted to an excitation signal $\bm{e}_{1: T}$, which is a sine-based signal for voiced segments and a DNN-transformed Gaussian white noise for unvoiced segments. By assuming the F0 value and V/UV flag value at the $t$-th time step are $f_t$ and $v_t$, the $t$-th value of excitation signal $e_t$ can be defined mathematically as
\begin{equation}
	e_t = 
	\begin{cases}
		\alpha \sin\Big(\sum\limits_{k=1}\limits^t 2\pi \frac{f_k}{N_s} + \phi\Big) + n_t, \quad v_t = 1\\
		g\Big(\frac{1}{3\sigma}n_t\Big), \qquad \qquad \qquad \qquad \quad v_t = 0 \quad
	\end{cases},
\end{equation}
where $v_t = 0$ or $1$ denotes that $e_t$ belongs to unvoiced or voiced segment, $\alpha$ and $\sigma$ are hyperparameters, $g(\cdot)$ represents a DNN-based transformation, $n_t \sim \mathcal{N}(0, \sigma^2)$ is a Gaussian noise, $\phi \in (-\pi, \pi]$ is a random initial phase, and $N_s$ denotes the waveform sampling rate. In the source module, only the DNN is trainable.

\subsection{Resolution-wise conditional filter module}

\begin{figure}[t!]
	\begin{center}
		\subfigure[SubBlock]{
			\centering
			\includegraphics[width=10cm]{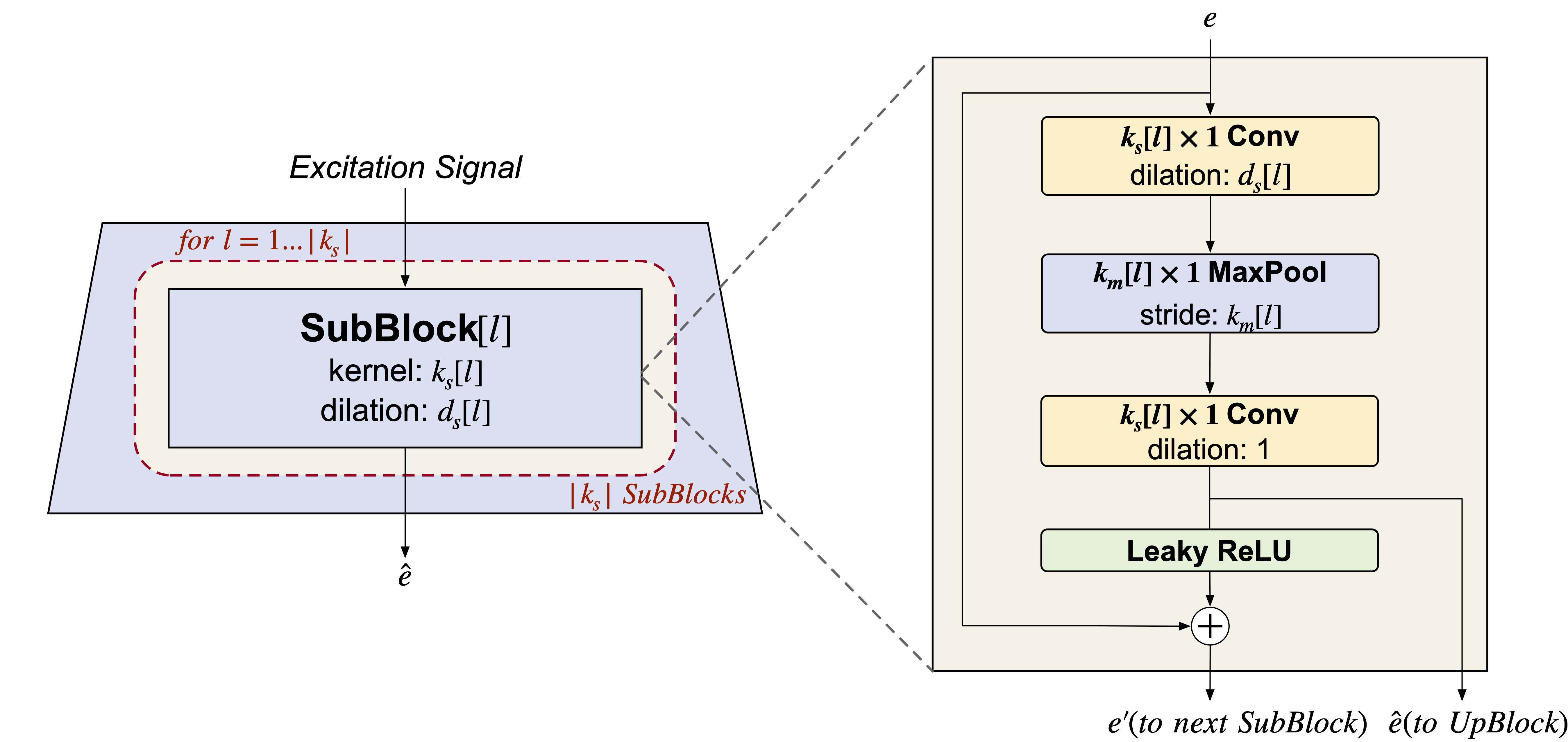}
			\label{fig:2 (a)}
        }
		\subfigure[UpBlock]{
			\centering
			\includegraphics[width=10cm]{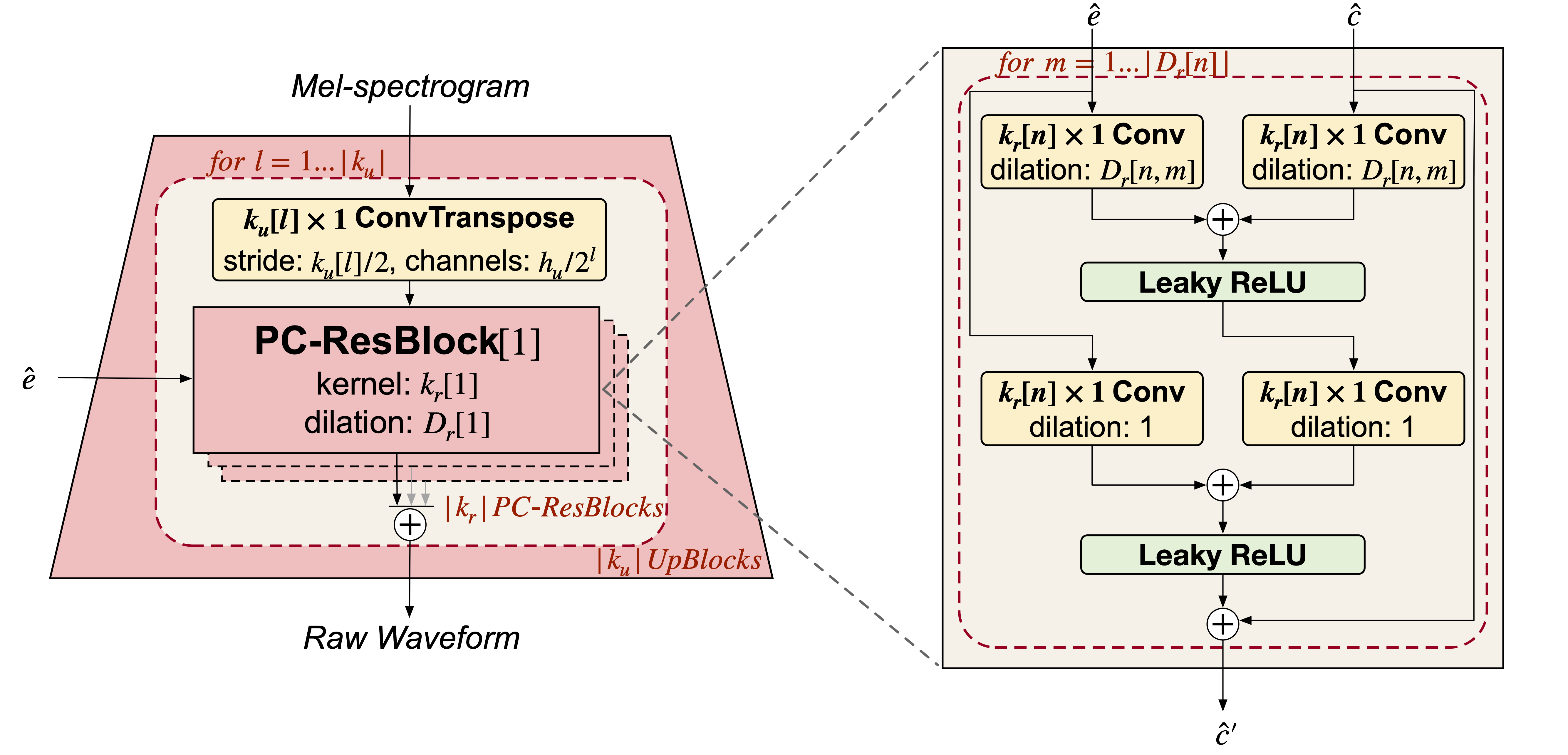}
			\label{fig:2 (b)}
		}
	\end{center}
	\caption{(a) The SubBlocks subsample the excitation signal $|k_s|$ times to match the resolution of intermediate features in each UpBlock with max-pooling. The $l$-th SubBlock with kernel size $k_s[l]$ and dilation rates $d_s[l]$ is depicted. (b) The UpBlocks upsample the mel-spectrogram $|k_u|$ times with the transposed convolutions and combine the upsampled features with the corresponding subsampled excitation signals in the PC-ResBlocks. The $n$-th PC-ResBlock with kernel size $k_r[n]$ and dilation rates $D_r[n]$ in the $l$-th UpBlock is depicted.}
	\label{fig:2}
\end{figure}

Given the mel-spectrogram and excitation signal as inputs, the filter module aims to reconstruct raw waveform from them. Our filter module is a combination of excitation subsampling blocks and the HiFi-GAN generator. The HiFi-GAN generator implements layer-wise upsampling on the frame-level mel-spectrogram to gradually match the temporal resolution of the speech waveform. In each upsampling layer, a transposed convolution is used for upsampling the intermediate feature, and an MRF module is used to observe patterns of various lengths in parallel. To condition the excitation signal to each layer of the HiFi-GAN generator, we propose the resolution-wise conditional filter (RCF) module. Specifically, inspired by the residual blocks (ResBlocks) in MRF, as shown in Fig.\ref{fig:2 (a)}, we design the subsampling blocks (SubBlocks), which apply layer-wise subsampling operations on the excitation signal to match the resolutions of corresponding intermediate features. And then, to combine the subsampled excitation signals with the corresponding intermediate features, we design parallel convolutional residual blocks (PC-ResBlocks) in the MRF module and conclude the redesigned MRF module together with the transposed convolution as an UpBlock. As illustrated in Fig.\ref{fig:2 (b)}, in each layer of the PC-ResBlock, a transformed excitation signal $\bm{\hat{e}}$ and an intermediate feature $\bm{\hat{c}}$ are converted to the next-level intermediate feature $\bm{\hat{c}'}$. To achieve this, we propose a feature to feature mapping 
\begin{equation}
		f_{k, d}(\bm{x}, \bm{y}) = LReLU(\bm{W}_{k, d} * \bm{x} + \bm{W}_{k, d} * \bm{y}),
\end{equation}
where $*$ denotes a convolution operator, $+$ denotes an element-wise addition operator, $LReLU$ is a leaky rectified linear unit (LReLU) \cite{maas2013rectifier} activation function, and $\bm{W}_{k, d}$ is a 1D-convolution with kernel size $k$ and dilation rate $d$. With this mapping function, the generative process of $\bm{\hat{c}'}$ can be defined as
\begin{equation}
	\bm{\hat{c}'} = f_{k, 1}(\bm{\hat{e}}, f_{k, d}(\bm{\hat{e}}, \bm{\hat{c}})) + \bm{\hat{c}}.
\end{equation}

Our proposed RCF module has the following advantages: Firstly, by feeding the excitation signal to a filter module, speech waveforms can be reconstructed under the direction of explicit fundamental frequencies for better frequency modeling. Secondly, with the resolution-wise condition architecture, which combines the input excitation signal and mel-spectrogram at different resolutions, our model can capture fundamental frequency information and acoustic properties across multiple frequency bands. Thirdly, through the MRF module with PC-Resblocks, our model can observe various receptive field patterns of excitation signals and acoustic features in parallel. In a way, the receptive fields are extended. 

\section{Experiments}

\subsection{Experimental setup}
To compare our model with others on seen data, we conducted experiments on the LJSpeech dataset \cite{ito2017lj}, which consists of 13,100 audio clips of a single English female speaker and is of about 24 hours. The audio sampling rate is 22.05 kHz with a format of 16-bit PCM. We randomly divided the dataset into training, validation, and test sets in a ratio of 8:1:1. Our proposed model was compared against the official implementation of HiFi-GAN \footnote{\url{https://github.com/jik876/hifi-gan}} and an open source implementation of Fre-GAN \footnote{\url{https://github.com/rishikksh20/Fre-GAN-pytorch}}. All the models were trained until 2.5M steps (about 3800 epochs).

To evaluate the generalization ability of our proposed model in a speaker-unseen scenario, we conducted multi-speaker experiments on the VCTK corpus \cite{veaux2017cstr}. The VCTK dataset consists of 44,257 audio clips from 109 native English speakers with various accounts, and the total length is about 44 hours. The audio sampling rate is 44 kHz with a format of 16-bit PCM, and we subsampled all the audio clips to 22.05 kHz. At the training stage, we randomly excluded nine speakers from the training set. In the rest 100 speakers, we randomly selected nine audio clips from each speaker for validation and trained the model with all the rest audio clips. At the generation stage, we randomly selected 100 audio clips from each unseen speaker to generate samples. We also trained our proposed SF-GAN together with HiFi-GAN and Fre-GAN until 2.5M steps.

Most of the above experiments were conducted based on two generator variations (i.e.,$V1$ and $V2$), while for the Fre-GAN, we just used its $V1$ version, which was adequate to reveal the performance differences between models. In the source module, we set $\alpha = 0.1$ and $\sigma = 0.003$. As the RCF module shown in Fig.~\ref{fig:2}, for $V1$, we set $h_u = 512$, $k_u = [16, 16, 4, 4]$, $k_r = [3, 7, 11]$, and $D_r = [[1, 3, 5] \times 3]$ for the UpBlocks while $k_m = [1, 2, 2, 8]$, $k_s = [15, 11, 7, 3]$, and $d_s = [7, 5, 3, 1]$ for the SubBlocks. The $V2$ version reduces the hidden dimensions $h_u$ to 128 but with the same receptive fields as the $V1$ version. We used F0 and 80-dimensional mel-spectrogram as input conditions, and both their Hanning window size and hop size were set to 1024 and 256, respectively. Additionally, for the mel-spectrogram, the FFT point number was set to 1024. At the training stage, we adopted the AdamW optimizer \cite{loshchilov2018decoupled}, and all the hyperparameters were set in accordance with HiFi-GAN. The synthesized audio samples are available at the demo website \footnote{\url{https://yxlu-0102.github.io/SF-GAN-Demo}}.

\begin{table*}[t!]
\centering\
\renewcommand{\arraystretch}{1.2}
\caption{Objective evaluation results for analysis-synthesis experiments.}
\begin{tabular}{llccccc}
\toprule[1pt]
\textbf{Dataset} & \textbf{Model} & \makecell[c]{\textbf{SNR}\\\textbf{(dB)}}$\uparrow$  & \makecell[c]{\textbf{LAS-RMSE}\\\textbf{(dB)}}$\downarrow$ & \makecell[c]{\textbf{MCD}\\\textbf{(dB)}}$\downarrow$ & \makecell[c]{\textbf{F0-RMSE}\\\textbf{(cent)}}$\downarrow$ & \makecell[c]{\textbf{V/UV error}\\\textbf{(\%)}}$\downarrow$  \\\hline
\multirow{5}{*}{LJSpeech} & HiFi-GAN $V1$ & 4.3186 & 6.3555 & 1.5340 & 42.3113 & 5.0804 \\
& Fre-GAN $V1$ & 4.4277& 6.2462& 1.5478& 40.3120 & 4.9076 \\
& SF-GAN $V1$ & \textbf{4.6161} & \textbf{6.1812} & \textbf{1.4229} & \textbf{33.7797} & \textbf{4.8340} \\\cline{2-7}
& HiFi-GAN $V2$ & 3.6501 & 6.9994 & 1.9805 & 48.3591 & 5.8613 \\
& SF-GAN $V2$ & \textbf{3.7774} & \textbf{6.6931} & \textbf{1.7629} & \textbf{44.9624} & \textbf{5.7700} \\\hline
\multirow{5}{*}{VCTK} & HiFi-GAN $V1$ & 2.1841 & 6.8098 & 2.3486 & 45.8586 & 8.3386 \\
& Fre-GAN $V1$ & 2.3693 & 6.5795 & 2.2559 & 38.7379 & 8.0487 \\
& SF-GAN $V1$ & \textbf{2.5788} & \textbf{6.4566} & \textbf{2.1518} & \textbf{34.8631} & \textbf{7.3406} \\\cline{2-7}
& HiFi-GAN $V2$ & 1.7438 & 7.3316 & 2.7787 & 48.0714 & 9.2443\\
& SF-GAN $V2$ & \textbf{1.9405} & \textbf{7.2284} & \textbf{2.5440} & \textbf{43.5458} & \textbf{8.1595} \\
\bottomrule[1pt]
\end{tabular}
\label{table:1}
\end{table*}

\begin{figure}[ht!]
	\centering
  \centerline{\includegraphics[width=\textwidth]{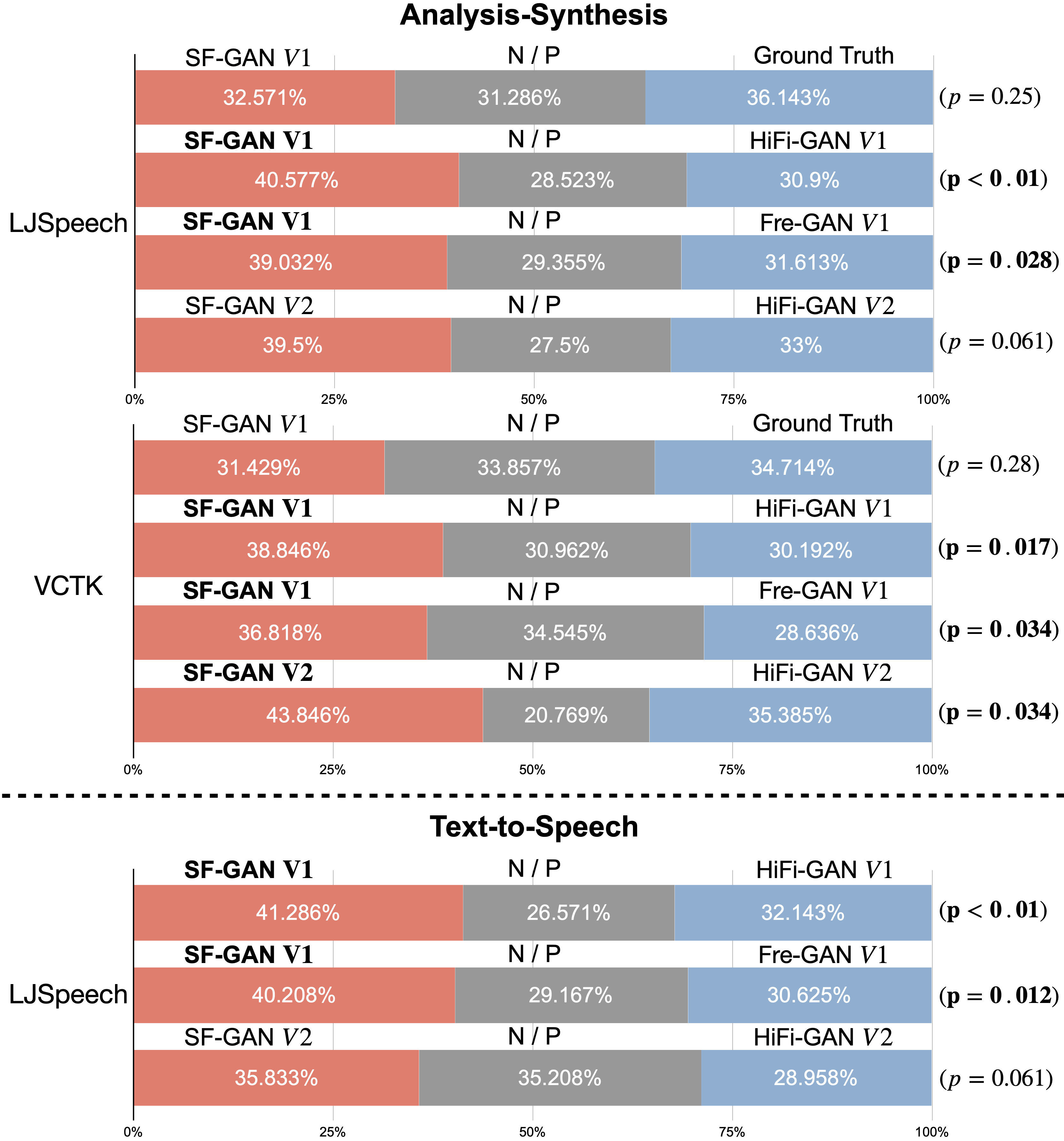}}
	\caption{Average preference scores ($\%$) of ABX tests on speech quality between SF-GAN and other three systems (i.e., Ground Truth, HiFi-GAN, and Fre-GAN), where N/P stands for ``no preference'' and $p$ denotes the $p$-value of a $t$-test between two systems.}
	\label{fig:3}
\end{figure}
 
\subsection{Comparison among neural vocoders}

\subsubsection{Evaluations on analysis-synthesis tasks.}

We implemented both objective and subjective evaluations on the LJSpeech and VCTK datasets to evaluate our proposed SF-GAN vocoder with HiFi-GAN and Fre-GAN in terms of synthesized speech quality. For objective evaluation, we adopted five metrics from our previous work \cite{ai2020neural}, including signal-to-noise ratio (SNR), root MSE (RMSE) of log amplitude spectra (LAS-RMSE), mel-cepstrum distortion (MCD), MSE of F0 (F0-RMSE), and V/UV error. For subjective evaluation, we conducted the ABX preference tests on the Amazon Mechanical Turk platform \footnote{\url{https://www.mturk.com}} to compare the differences between two comparative systems. In each ABX test, 20 utterances synthesized by two comparative systems were randomly selected from the test set and evaluated by at least 30 native English listeners. The listeners were asked to judge which utterance in each pair had better speech quality or whether there was no preference. In order to calculate the average preference scores, the $p$-value of a $t$-test was used to measure the significance of the difference between each two systems.

For experiments on the LJSpeech dataset, the objective results are presented in the top half of Table \ref{table:1}. Obviously, our proposed SF-GAN vocoder outperformed HiFi-GAN and Fre-GAN among all the objective metrics, demonstrating the distinct advantages of the proposed model in synthesized speech quality. The subjective ABX test results are illustrated in the top part of Fig.\ref{fig:3}, which shows that the performance of our proposed SF-GAN $V1$ was comparable with the ground-truth natural speech ($p = 0.25$) and significantly better than HiFi-GAN $V1$ ($p < 0.01$) and Fre-GAN $V1$ ($p < 0.05$). In addition, the performance of SF-GAN $V2$ was slightly better than HiFi-GAN $V2$ ($p$ is slightly higher than 0.05). These above experimental results verified the advantage of our model in improving synthesized speech quality.

For experiments on the VCTK datasets, we used utterances of nine unseen speakers excluded from the training set for the objective evaluation and randomly selected 20 utterances from them for the ABX test as above. The objective results are presented in the bottom half of Table \ref{table:1}, and among all the metrics, our proposed SF-GAN surpassed HiFi-GAN and Fre-GAN. The subjective ABX test results are illustrated in the middle part of Fig.\ref{fig:3}, which also shows that the performance of our proposed SF-GAN was better than that of HiFi-GAN in both $V1$ and $V2$ ($p < 0.05$) and Fre-GAN in $V1$ ($p < 0.05$). Moreover, it is noteworthy that even with unseen speakers, the performance of SF-GAN was still comparable with ground-truth audios ($p = 0.28$), which verified the generalization ability of our proposed model in the speaker-unseen scenario.

\begin{figure}[t!]
	\centering
  \centerline{\includegraphics[width=6cm]{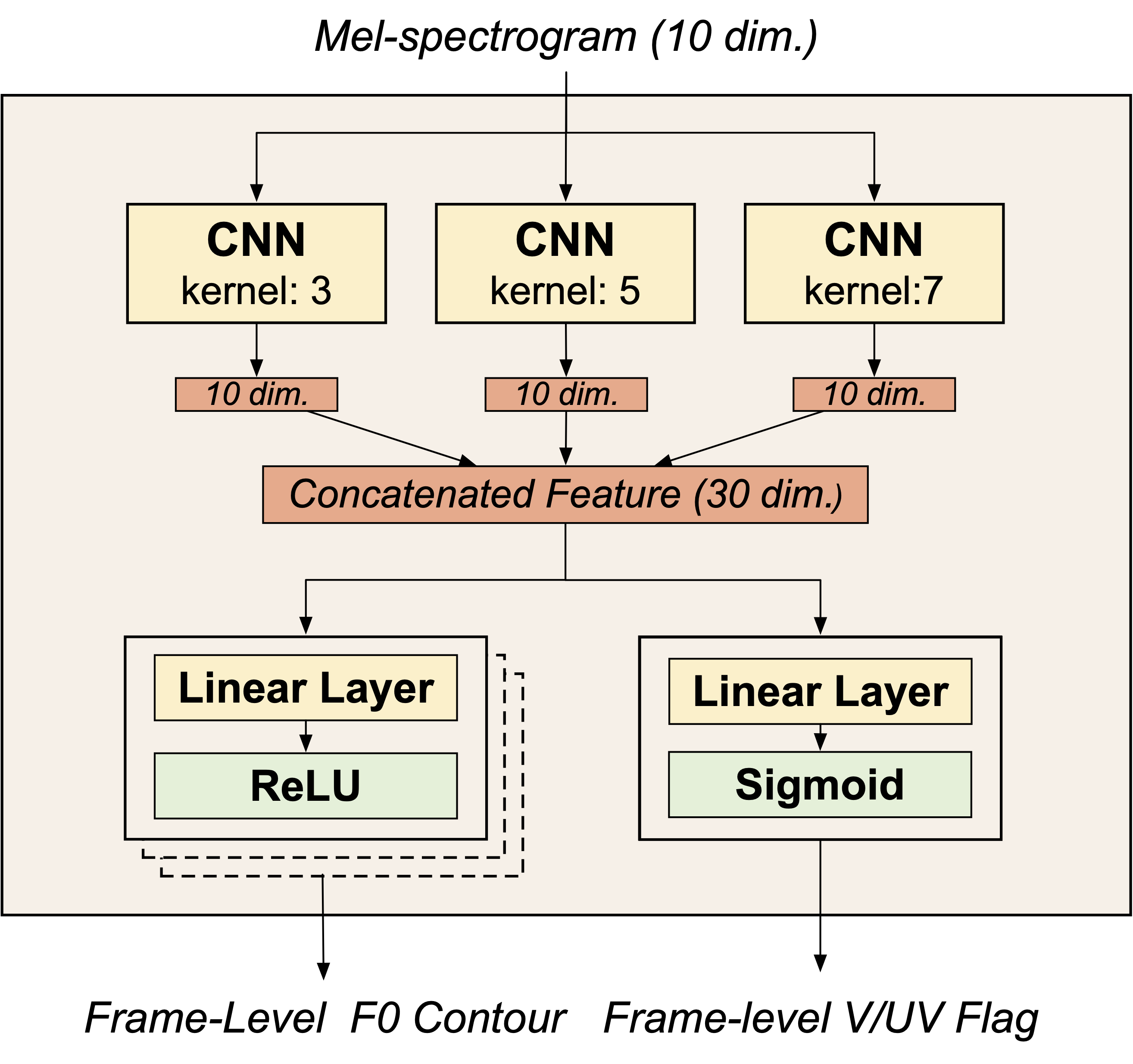}}
	\caption{The architecture of the F0 predictor. We use predicted F0 and synthesized mel-spectrogram as the inputs of the F0 predictor in the TTS task.}
	\label{fig:4}
\end{figure}

\begin{figure}[t!]
	\centering
  \centerline{\includegraphics[width=\textwidth]{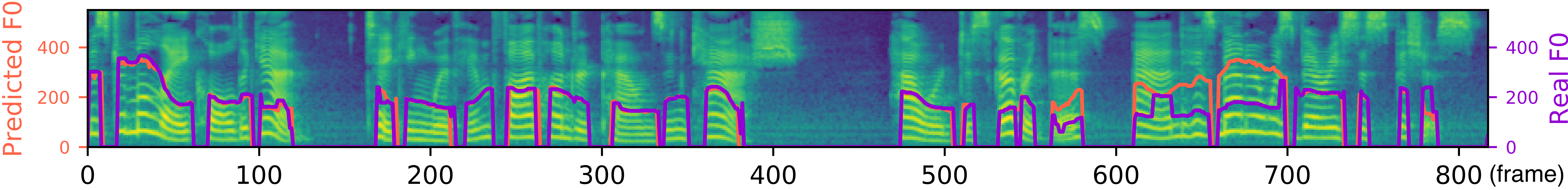}}
	\caption{An example of comparison between a predicted F0 and a real F0. The red curve denotes the predicted F0 and the purple curve denotes the real F0.}
	\label{fig:5}
\end{figure}

\subsubsection{Evaluations on TTS tasks.}
Since our proposed SF-GAN needs an additional F0 input, which is distinct from other mel-spectrogram vocoders.
To apply our proposed model to the TTS task, we proposed an F0 predictor to predict F0 from the first ten dimensions of the mel-spectrogram, which contains sufficient F0 information. The architecture of the F0 predictor is illustrated in Fig.\ref{fig:4}. We first input the first 10-dimensional mel-spectrogram into three convolutional neural networks (CNNs) with different kernel sizes and get three 10-dimensional intermediate features. We then input the concatenated 30-dimensional feature to two linear layers with different activation functions (i.e., ReLU and Sigmoid) to get the F0 contour and V/UV flag, respectively. We used the LJSpeech dataset to train and test the F0 predictor with the same data division as in the analysis-synthesis experiment, and we evaluated the synthesized F0 contours and V/UV flags from real mel-spectrograms with two objective metrics, including F0-RMSE and V/UV error. The F0-RMSE and V/UV errors on the test set are 115.7072 cents and 4.3901\%, respectively. A close examination found that most of the errors came from the inaccuracy of F0 extraction. An example is shown in Fig.~\ref{fig:5}, where we can see the F0 extraction errors occurred between the 560-th frame and the 700-th frame.

\begin{figure}[t!]
	\begin{center}
		\subfigure[SF-GAN]{
			\centering
			\includegraphics[width=10cm]{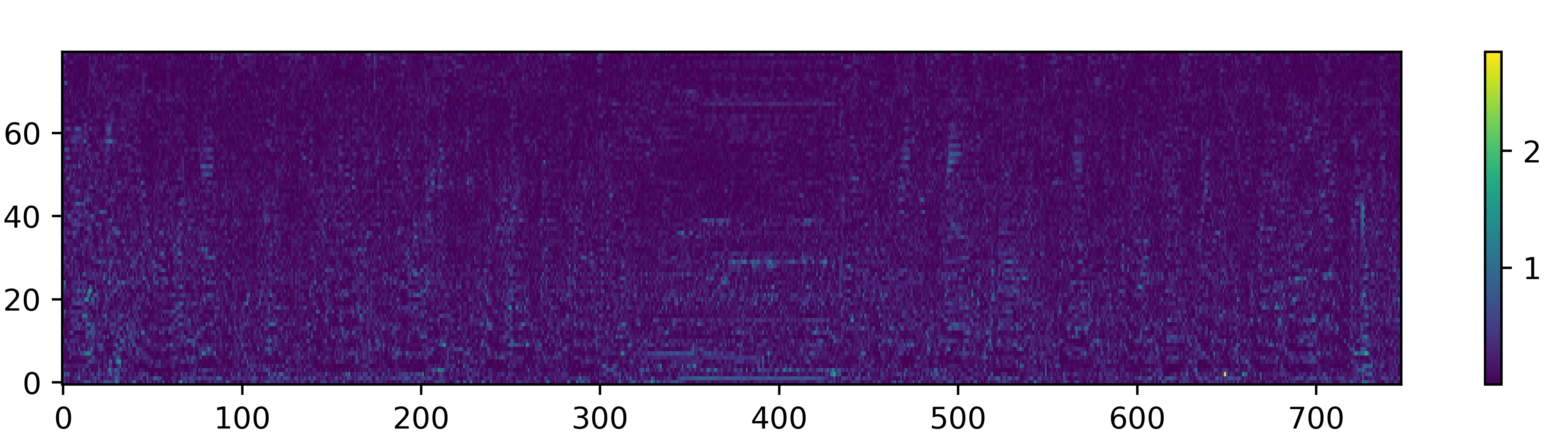}
			\label{fig:6 (a)} 
		}
		\subfigure[Fre-GAN]{
			\centering
			\includegraphics[width=10cm]{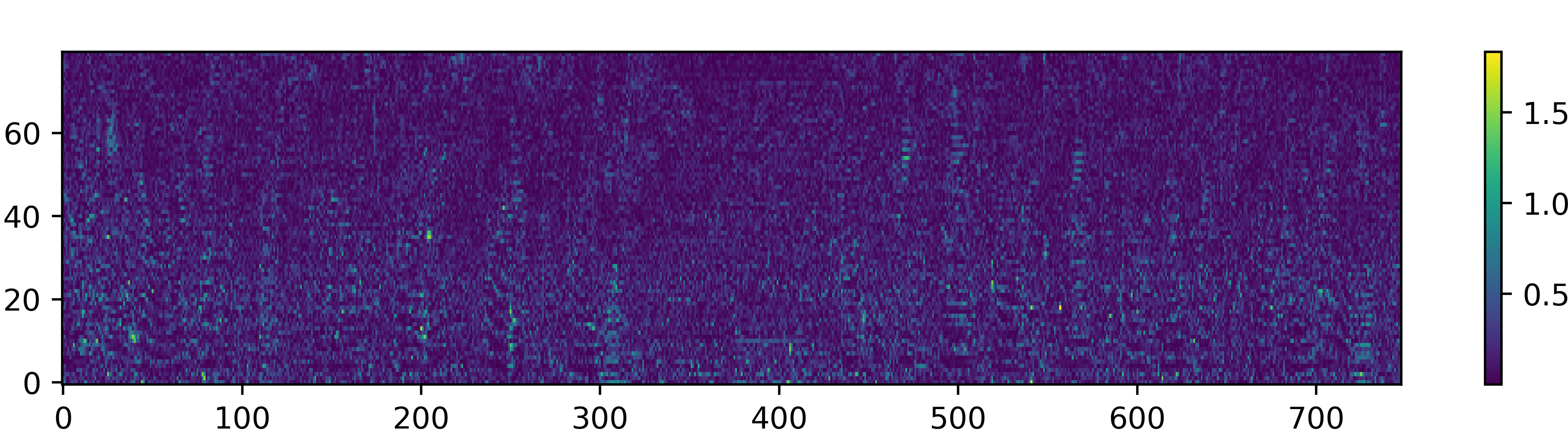}
			\label{fig:6 (a)} 
		}
		\subfigure[HiFi-GAN]{
			\centering
			\includegraphics[width=10cm]{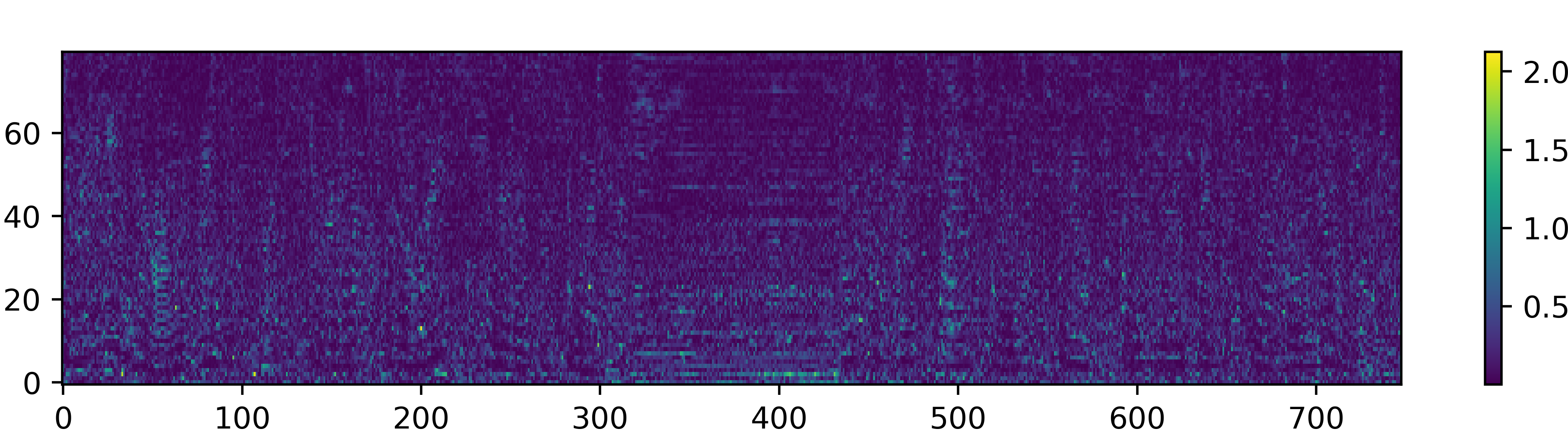}
			\label{fig:6 (b)}
		}
	\end{center}
	\caption{Pixel-wise difference between a mel-spectrogram from Tacotron2 and mel-spectrograms from waveforms generated by SF-GAN, Fre-GAN, and HiFi-GAN.}
	\label{fig:6}
\end{figure}

We herein predicted the mel-spectrograms from texts by adopting the most popular implementation of Tacotron2 \footnote{\url{https://github.com/NVIDIA/tacotron2}} \cite{shen2018natural} with the provided pre-trained weights on the LJSpeech dataset. We fed the mel-spectrogram predicted by Tacotron2 as the input condition to HiFi-GAN and Fre-GAN. Then we predicted the F0 contour and V/UV flag from the predicted mel-spectrogram and fed all of them as input conditions to our proposed SF-GAN. We also conducted ABX preference tests between them and the results are illustrated in the bottom part of Fig.~\ref{fig:3}, which demonstrates our proposed SF-GAN $V1$ significantly outperformed HiFi-GAN $V1$ ($p < 0.01$) and Fre-GAN $V1$ ($p < 0.05$), while SF-GAN $V2$ was slightly better than HiFi-GAN $V2$ ($p$ is slightly higher than 0.05).
In addition, when we investigated the pixel-wise mel-spectrogram difference between a generated mel-spectrogram from Tacotron2 and extracted mel-spectrograms from waveforms generated by SF-GAN, HiFiGAN, and Fre-GAN, as presented in Fig.~\ref{fig:6}, it's apparent that the pixel-wise difference of our model is less than those of HiFi-GAN and Fre-GAN. To sum up, we conclude that our SF-GAN can be well applied to the TTS task. 

\begin{table}[t!]
\centering\
\renewcommand{\arraystretch}{1.2}
\caption{Objective evaluation results for ablation studies.}
\begin{tabular}{lccccc}
\toprule[1pt]
\textbf{Model} & \makecell[c]{\textbf{SNR}\\\textbf{(dB)}}$\uparrow$  & \makecell[c]{\textbf{LAS-RMSE}\\\textbf{(dB)}}$\downarrow$ & \makecell[c]{\textbf{MCD}\\\textbf{(dB)}}$\downarrow$ & \makecell[c]{\textbf{F0-RMSE}\\\textbf{(cent)}}$\downarrow$ & \makecell[c]{\textbf{V/UV error}\\\textbf{(\%)}}$\downarrow$  \\\hline
SF-GAN $V2$ & 3.7774 & \textbf{6.6931} & \textbf{1.7629} & 44.9624 & \textbf{5.6931} \\\cline{1-6}
w/o DNN & 3.8773 & 6.7600 & 1.8197 & \textbf{42.7686} & 5.7700\\
w/o SubBlock & 3.7111& 6.9126 & 1.9078 & 47.1645 & 6.0300 \\
w/o PC-ResBlock & \textbf{3.8921} & 6.8701 & 1.8472 & 44.7955 & 5.7856 \\\cline{1-6}
HiFi-GAN $V2$ & 3.6501 & 6.9994 & 1.9805 & 48.3591 & 5.8613 \\
\bottomrule[1pt]
\end{tabular}
\label{table:2}
\end{table}

\begin{figure}[t!]
	\centering
  \centerline{\includegraphics[width=\textwidth]{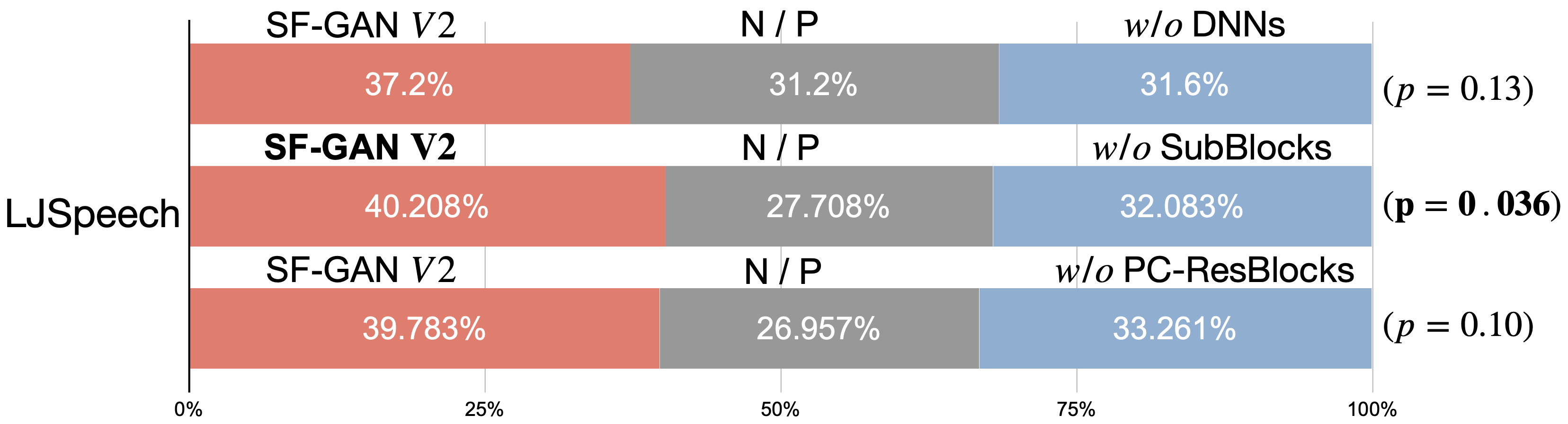}}
	\caption{The average preference scores ($\%$) of ABX tests between SF-GAN $V2$ and its ablation models in speech quality, where N/P stands for ``no preference'' and $p$ denotes the $p$-value of a $t$-test between two systems.}
	\label{fig:7}
\end{figure}

\subsection{Ablation studies}

We implemented ablation studies on the  DNN, SubBlock, and PC-ResBlock used in the SF-GAN vocoder to verify the effectiveness of each component in terms of synthesized speech quality in the analysis-synthesis task, where ablating DNN denotes removing the DNN in the source module, ablating SubBlock denotes directly subsampling the excitation signal to four resolutions instead of using SubBlocks, and ablating PC-ResBlock denotes simply adding the subsampled excitation signals with the upsampled intermediate features followed by the ResBlocks instead of combining them in the PC-ResBlocks. The $V2$ version with fewer hidden dimensions was used as the generator for the ablation studies, and all the ablation models were trained until 2.5M steps. 

As objective results presented in Table~\ref{table:2}, ablating SubBlock demonstrates distinct degradation among all the metrics while ablating DNN and PC-ResBlock show a slight degradation in partial metrics. To further verify the effect of these components subjectively, we also conducted ABX tests. The subjective results are illustrated in Fig.~\ref{fig:7}, which demonstrates that all three components contribute to the improvement. In accordance with the objective results, SubBlock remarkably contributes to the synthesized speech quality ($p < 0.05$), while ablating DNN ($p = 0.13$) or PC-ResBlock ($p = 0.10$) cause slight but insignificant quality degradations.

\label{sec:experiment}

\section{Conclusion}

In this work, we proposed the SF-GAN vocoder, which can synthesize high-quality speech from input F0 and mel-spectrogram. We took inspiration from the layer-wise upsampling architecture of HiFi-GAN and applied a resolution-wise source-filter access method to the HiFi-GAN framework where excitation signal generated by a source module and mel-spectrogram are combined at various resolutions, which significantly contributes to the quality of synthesized speech audios according to the ablation studies. Above all, the proposed SF-GAN vocoder outperforms the state-of-the-art HiFi-GAN and Fre-GAN in both analysis-synthesis and TTS tasks of speech synthesis, and the speech quality of synthesized audios by SF-GAN is even comparable with that of natural ones. It is noteworthy that the SF-GAN vocoder shows strong generalization ability in the speaker-unseen scenario and the ability to be applied to the TTS task. In our future work, we will apply our proposed method to other GAN-based vocoders.

\label{sec:conclusion}

\subsubsection{Acknowledgment.}

This work was partially funded by the National Nature Science Foundation of China under Grant 61871358.

%
%
%
\bibliographystyle{splncs04_unsort}
\bibliography{mybibliography}

\begin{thebibliography}{10}
\providecommand{\url}[1]{\texttt{#1}}
\providecommand{\urlprefix}{URL }
\providecommand{\doi}[1]{https://doi.org/#1}

\bibitem{zen2009statistical}
Zen, H., Tokuda, K., Black, A.W.: Statistical parametric speech synthesis.
  speech communication  \textbf{51}(11),  1039--1064 (2009)

\bibitem{kawahara1999restructuring}
Kawahara, H., Masuda-Katsuse, I., De~Cheveigne, A.: Restructuring speech
  representations using a pitch-adaptive time--frequency smoothing and an
  instantaneous-frequency-based f0 extraction: Possible role of a repetitive
  structure in sounds. Speech communication  \textbf{27}(3-4),  187--207 (1999)

\bibitem{morise2016world}
Morise, M., Yokomori, F., Ozawa, K.: {WORLD}: A vocoder-based high-quality
  speech synthesis system for real-time applications. IEICE TRANSACTIONS on
  Information and Systems  \textbf{99}(7),  1877--1884 (2016)

\bibitem{oord2016wavenet}
Oord, A.v.d., Dieleman, S., Zen, H., Simonyan, K., Vinyals, O., Graves, A.,
  Kalchbrenner, N., Senior, A., Kavukcuoglu, K.: Wave{N}et: A generative model
  for raw audio. In: 9th ISCA Speech Synthesis Workshop. pp. 125--125 (2016)

\bibitem{mehri2016samplernn}
Mehri, S., Kumar, K., Gulrajani, I., Kumar, R., Jain, S., Sotelo, J.,
  Courville, A., Bengio, Y.: Sample{RNN}: An unconditional end-to-end neural
  audio generation model. In: Proc. ICLR (2017)

\bibitem{kalchbrenner2018efficient}
Kalchbrenner, N., Elsen, E., Simonyan, K., Noury, S., Casagrande, N., Lockhart,
  E., Stimberg, F., Oord, A., Dieleman, S., Kavukcuoglu, K.: Efficient neural
  audio synthesis. In: Proc. ICML. pp. 2410--2419 (2018)

\bibitem{oord2017parallel}
Oord, A.v.d., Li, Y., Babuschkin, I., Simonyan, K., Vinyals, O., Kavukcuoglu,
  K., Driessche, G.v.d., Lockhart, E., Cobo, L.C., Stimberg, F., et~al.:
  Parallel {W}ave{N}et: Fast high-fidelity speech synthesis. In: Proc. ICML.
  pp. 3918--3926 (2018)

\bibitem{ping2018clarinet}
Ping, W., Peng, K., Chen, J.: Clari{N}et: Parallel wave generation in
  end-to-end text-to-speech. In: Proc. ICLR (2019)

\bibitem{prenger2018waveglow}
Prenger, R., Valle, R., Catanzaro, B.: Wave{G}low: A flow-based generative
  network for speech synthesis. In: Proc. ICASSP. pp. 3617--3621 (2019)

\bibitem{ping2020waveflow}
Ping, W., Peng, K., Zhao, K., Song, Z.: Wave{F}low: A compact flow-based model
  for raw audio. In: Proc. ICML. pp. 7706--7716 (2020)

\bibitem{wang2019neural}
Wang, X., Takaki, S., Yamagishi, J.: Neural source-filter-based waveform model
  for statistical parametric speech synthesis. In: Proc. ICASSP. pp. 5916--5920
  (2019)

\bibitem{goodfellow2014generative}
Goodfellow, I., Pouget-Abadie, J., Mirza, M., Xu, B., Warde-Farley, D., Ozair,
  S., Courville, A., Bengio, Y.: Generative adversarial nets. In: Proc.
  NeurIPS. pp. 2672--2680 (2014)

\bibitem{donahue2018adversarial}
Donahue, C., McAuley, J., Puckette, M.: Adversarial audio synthesis. In: Proc.
  ICLR (2018)

\bibitem{binkowski2019high}
Bi{\'n}kowski, M., Donahue, J., Dieleman, S., Clark, A., Elsen, E., Casagrande,
  N., Cobo, L.C., Simonyan, K.: High fidelity speech synthesis with adversarial
  networks. In: Proc. ICLR (2019)

\bibitem{kumar2019melgan}
Kumar, K., Kumar, R., de~Boissiere, T., Gestin, L., Teoh, W.Z., Sotelo, J.,
  de~Br{\'e}bisson, A., Bengio, Y., Courville, A.C.: {MelGAN}: Generative
  adversarial networks for conditional waveform synthesis. Advances in neural
  information processing systems  \textbf{32} (2019)

\bibitem{yamamoto2020parallel}
Yamamoto, R., Song, E., Kim, J.M.: Parallel {W}ave{GAN}: A fast waveform
  generation model based on generative adversarial networks with
  multi-resolution spectrogram. In: Proc. ICASSP. pp. 6199--6203 (2020)

\bibitem{kong2020hifi}
Kong, J., Kim, J., Bae, J.: {HiFi-GAN}: Generative adversarial networks for
  efficient and high fidelity speech synthesis. Advances in Neural Information
  Processing Systems  \textbf{33},  17022--17033 (2020)

\bibitem{kim2021fre}
Kim, J.H., Lee, S.H., Lee, J.H., Lee, S.W.: {Fre-GAN}: Adversarial
  frequency-consistent audio synthesis. In: Proc. InterSpeech 2021. pp.
  3246--3250 (2021)

\bibitem{wu2021quasi}
Wu, Y.C., Hayashi, T., Okamoto, T., Kawai, H., Toda, T.: {Quasi-periodic
  Parallel WaveGAN}: A non-autoregressive raw waveform generative model with
  pitch-dependent dilated convolution neural network. IEEE/ACM Transactions on
  Audio, Speech, and Language Processing  \textbf{29},  792--806 (2021)

\bibitem{yoneyama2021unified}
Yoneyama, R., Wu, Y.C., Toda, T.: Unified source-filter {GAN}: Unified
  source-filter network based on factorization of quasi-periodic parallel
  wavegan. In: Proc. InterSpeech 2021. pp. 2187--2191 (2021)

\bibitem{yoneyama2022unified}
Yoneyama, R., Wu, Y.C., Toda, T.: Unified source-filter {GAN} with
  harmonic-plus-noise source excitation generation. arXiv preprint
  arXiv:2205.06053  (2022)

\bibitem{maas2013rectifier}
Maas, A.L., Hannun, A.Y., Ng, A.Y.: Rectifier nonlinearities improve neural
  network acoustic models. In: Proc. ICML. vol.~30, p.~3 (2013)

\bibitem{ito2017lj}
Ito, K., Johnson, L.: {The lj speech dataset}.
  \url{https://keithito.com/LJ-Speech-Dataset/} (2017)

\bibitem{veaux2017cstr}
Veaux, C., Yamagishi, J., MacDonald, K., et~al.: {CSTR VCTK corpus}: English
  multi-speaker corpus for {CSTR} voice cloning toolkit. University of
  Edinburgh. The Centre for Speech Technology Research (CSTR)  (2017)

\bibitem{loshchilov2018decoupled}
Loshchilov, I., Hutter, F.: Decoupled weight decay regularization. In:
  International Conference on Learning Representations (2018)

\bibitem{ai2020neural}
Ai, Y., Ling, Z.H.: A neural vocoder with hierarchical generation of amplitude
  and phase spectra for statistical parametric speech synthesis. IEEE/ACM
  Transactions on Audio, Speech, and Language Processing  \textbf{28},
  839--851 (2020)

\bibitem{shen2018natural}
Shen, J., Pang, R., Weiss, R.J., Schuster, M., Jaitly, N., Yang, Z., Chen, Z.,
  Zhang, Y., Wang, Y., Skerrv-Ryan, R., et~al.: Natural tts synthesis by
  conditioning wavenet on mel spectrogram predictions. In: 2018 IEEE
  international conference on acoustics, speech and signal processing (ICASSP).
  pp. 4779--4783. IEEE (2018)

\end{thebibliography}
%

%
%
%

\end{document}